\documentclass[reqno]{amsart}
\usepackage{graphicx,amsmath,color}
\usepackage[T1]{fontenc}
\usepackage[latin2]{inputenc}

\usepackage{latexsym,amsfonts,amssymb,amsmath,amsthm}
\usepackage{paralist}
\usepackage{fancyhdr}
\usepackage{lmodern}

\DeclareMathAlphabet{\mathbk}{OT1}{cmr}{bx}{sl}

\newcommand\uu{\mathbk u} \newcommand\uc{\uu_c}
\newcommand\nn{\mathbk n} \newcommand\nc{\nn_c}
\newcommand\ep{\epsilon} \newcommand\en{_{\ep\nn}}
\renewcommand\sc{s_c} \newcommand\sen{s\en}
\newcommand\aen{a\en} \newcommand\ren{r\en}
\newcommand\ac{a_c} 
\newcommand\beq{\begin{equation}}
\newcommand\enq{\end{equation}}
\newcommand\ran{\mathrm{Ran}}

\newcommand\1{\boldsymbol 1}
\newcommand\Dd{\mathrm D}
\newcommand\z{\mathbk z} \newcommand\h{\mathbk h}
\newcommand\x{\mathbk x} \newcommand\y{\mathbk y}

\newcommand\aM{\mathrm M}\newcommand\M{\mathbf M}
\newcommand\la{\lambda}
\newcommand\rr{\mathbb R}

\newcommand\jj{\mathbk j}
\newcommand\Ll{\mathrm L^\rightarrow}
\newcommand\q{\mathbk q}
\newcommand\ad{\mathbk a}\newcommand\bd{\mathbk b}
\newcommand\Lt{\mathbk L}
\newcommand\I{\mathbb T}

\newcommand\E{\mathbf E}
\newcommand\Pp{\mathbk P}
\newcommand\ordo{\text{ordo}}
\newcommand\Sd{\mathbf S}
\newcommand\Dt{\mathcal D}
\newcommand\Tr{\mathrm{Tr}}

\begin{document}

\title[Radiation reaction]{Testing radiation reaction by simple charge distributions}


\author{T. Matolcsi$^1$ and P. V\'an$^{2,3,4}$ }
\address{
	$^1$ Department of Applied Analysis and Computational Mathematics, E\"otv\"os Lor\'and University, Budapest, Hungary;//
	$^2$ Department of Theoretical Physics, Wigner Research Centre for Physics, H-1525 Budapest, Konkoly Thege MiklĂłs u. 29-33., Hungary; //
	and  
	$^3$Department of Energy Engineering, Faculty of Mechanical Engineering,  Budapest University of Technology and Economics, 1111 Budapest, MĹąegyetem rkp. 3., Hungary,//
	$^4$Montavid Thermodynamic Research Group, Budapest, Hungary.}

\begin{abstract}
A point charge is frequently approximated by various charge distributions deriving the Lorentz-Abraham-Dirac (LAD) equation. Here a rigid spherical shell is treated from this point of view. This particular continuum model is excellent to investigate whether some hidden expectations are true or not. It is shown here that the field of a uniformly charged rigid spherical shell cannot be substituted by the field of a central point charge.The related calculations are short and transparent due to a reference frame free representation of spacetime and the mathematical theory of generalised functions. 
\end{abstract}
\maketitle

\section{Introduction}

The Lorentz-Abraham-Dirac(LAD) equation, \cite{Lor892a,Abr905b,Lor09b,Dir38a}, is the simplest possible model of field-matter interaction, coupling Maxwell equations to Newton equation, electrodynamics to mechanics and mechanics to electrodynamics. LAD equation, however, is a wrong model of any phenomena, because of  unacceptable mathematical and physical properties. There are numerous attempts to remedy the related problems, with various and different strategies to remove the mathematical and physical inconsistencies. These strategies are the following:
\begin{itemize}
\item Nonlocal modifications of the electromagnetic part, the Maxwell equations, \cite{BorInf34a,Bop40a,Kie19a}.
\item Dissipative modifications of the mechanical part, the Newton equation \cite{ForOco91a,Ver08a,Pol14a,ForEta20a}.
\item A suitable interpretation of the LAD equation, e.g. excluding particular solutions, \cite{Dir38a,Spo00a,MatEta17a}.
\item Application of continuum charge distributions instead of point charges. This is the method of the classical papers of Lorentz, Abraham and Dirac, too, \cite{Lor892a,Abr905b,Lor09b,Dir38a}. There are two main aspects of this strategy: 
	\begin{itemize} 
		\item One may improve the classical theory, with the identification and elimination of the mathematical problems, \cite{Par87b,BilEta19a},
		\item One may modify the point charge model with the help of quantum mechanics, or with various renormalization procedures, \cite{MonSha74a,MonSha77a,GalEta10a,ForEta12c,GalEta12c,Kos19a,Pol19a}.
	\end{itemize}
\end{itemize}

The related literature is vast, the interested reader may look into the monographs of the field, \cite{Spo04b,Yag06b,Roh07b,Kos07b}. Worth mentioning, that recent experiments of radiation reaction related phenomena, see, e.g. \cite{WisEta18a,PodEta18a,Bla20a}, open the way toward the verification of the mentioned theories and also toward less rigorous, but applicable approaches, \cite{LanLif78b,Sok09a,Zot16a}.

The mathematical side of the problem is that the Maxwell equations are singular along the worldline of a point particle. A related issue is the infinite electromagnetic energy of a point charge or a line current. That is why continuum charge distributions are promising, where one derives the corresponding point charge equation of motion by a limiting procedure \cite{Ger96m,GraEta09a}. Here the simplest possible model is probably a rigid, uniformly charged spherical shell. However, it is not simple, various pitfalls of the related calculations were discovered step by step, \cite{Dir38a,Par87b}. 

The theory of generalised functions, \cite{Dij13b}, is the mathematical tool to deal with simple singularities. It is easy to show, that electromagnetic energies of a point charge or a line current are not singular at all if they are represented by distributions \cite{Mat20bh}. Moreover, in this way we can obtain the radiation reaction force without approximations, limiting procedures and subtraction of infinite quantities \cite{Mat20m}.

Recently  Bild, Decker and Ruhl \cite{BilEta19a} analysed the problem of the LAD equation with the help of a continuous charge distribution. They pointed to that the equation of motion for a point charge obtained from truncated series requires the proof of the convergence. Moreover, they have observed that the definition of the world tube and the Gauss-Stokes-Ostrogradski theorem must be correctly applied, when contrasted to the derivation of Dirac \cite{Dir38a}. They have obtained a delayed equation of motion, with promising properties, like it was obtained previously by Raju and Raju in a more intuitive manner \cite{RajRaj08a}. 

In this work we investigate a simple case whether a homogeneous charge distribution of a spherical shell could be replaced by an effective charge at the centre of the sphere. In case of inertial motion this is a trivial consequence of the Maxwell equations, we will show that for a uniformly accelerated sphere it is not the case.


Our calculation is based on distribution theory, and a reference frame free model of special relativistic spacetime ensures the precise mathematical treatment. 
considerably. The necessary mathematical concepts are collected in the Appendix.

\section{About our formalism}

{A condensed discussion is possible using a coordinate and reference-free treatment of spacetime which can be found in the book \cite{Mat20b}. }

A brief summary of that (with the speed of the light $c=1$):

Spacetime points and spacetime vectors (which are often confused in coordi\-nates) are distinguished:

\begin{itemize}
	\item $\aM$, mathematically a four dimensional affine space, is the set of {\em spacetime points}, 
	\item $\M$,  mathematically a four dimensional vector space, is the set of {\em  spacetime vectors},
	\item the set of {\em time periods} is $\I$, a one dimensional oriented vector space,
	\item the exact treatment requires the tensorial quotients of $\M$ by $\I$, here we do not refer to them explicitely, 
	\item the Lorentz product of the spacetime vectors $\x$ and $\y$ is denoted by $\x\cdot\y$ (which is $x_ky^k$ in coordinates);
$\x$ is timelike if $\x\cdot\x<0$ (which means in coordinates the signature $(-1,1,1,1)$ of the Lorentz form),
	\item an {\em absolute velocity} is a futurelike vector $\uu$ for which $\uu\cdot\uu=-1$,
	\item for an absolute velocity  $\uu$, $\Sd_\uu:=\{\x\mid \uu\cdot\x=0\}$ is the set of {\em $\uu$-spacelike vectors}, a three dimensional Euclidean subspace of $\M$
	\item the action of linear and bilinear maps is denoted by a dot, too; e.g the action of a linear map $\Lt$ on a vector 
$\x$ is $\Lt\cdot\x$ (which is ${L^i}_kx^k$ in coordinates),
	\item the adjoint of a linear map $\Lt$ is the linear map $\Lt^*$ defined by $(\Lt^*\cdot\x)\cdot\y=\x\cdot\Lt\cdot\y$ for all vectors
$\x,\y$; shortly, $\Lt^*\cdot\x=\x\cdot\Lt$; similarly, $\y\cdot\Lt^*=\Lt\cdot\y$,
	\item $\Lt$ is a {\em  Lorentz transformation} if and only if $\Lt^*=\Lt^{-1}$,
	\item $\1$ denotes the identity map of vectors (which is ${g^i}_k={\delta^i}_k$ in coordinates).
\end{itemize}
Some other required consequent mathematical background is given in the Appendix.

\section{Uniformly accelerated sphere -- a world tube}

Let us consider the rigid uniformly accelerated observer treated in subsection II.6.4 of \cite{Mat20b}. Its space points are uniformly accelerated world lines.

Pick up such a world line, as the {\em center of the sphere} being the range of the world line function,
$$
	r_c(\sc)=x_c + \uc\frac{\sinh(\ac\sc)}{\ac} + \nc\frac{\cosh(\ac\sc)-1}{\ac}.
$$
where 
\begin{itemize}
	\item $x_c$ is an arbitrarily chosen spacetime point of the world line, 
	\item $\sc$ is the proper time of the world line,
	\item $\ac$ is the magnitude of the acceleration of the world line,
	\item $\uc$ is an absolute velocity and $\nc$ is an $\uc$-spacelike unit vector, that is $\uc\cdot \nc = 0$ and  $\nc\cdot \nc = 1$
\end{itemize}

The corresponding absolute velocity function is
\beq\label{dotrc}
	\dot r_c(\sc)=\uc\cosh(\ac\sc) + \nc\sinh(\ac\sc).
\enq

Further, we take other space points (world lines in spacetime) of the observer in question in order to represent a {\em uniformly accelerated sphere} as follows.

Let us take the set of $\uc$-spacelike unit vectors,
\beq\label{spherc}
	S_c(1):=\{\nn\mid \uc\cdot\nn=0, \ \nn\cdot\nn=1\};
\enq
for all of its elements and for an $\ep>0$ we consider the uniformly accelerated world line function
$$
	\ren(\sen)=x_c + \ep\nn + \uc\frac{\sinh(\aen\sen)}{\aen} + \nc\frac{\cosh(\aen\sen)-1}{\aen}.
$$
for which
\beq
	\dot r\en(\sen)=\uc\cosh(\aen\sen) + \nc\sinh(\aen\sen)
\enq
where, of course, $\dot{ }$ denotes the differentiation by the proper time $\sen$. 

The observer applies the synchronization according to which the proper times of
$r_c$ and $\ren$ are simultaneous if and only if 
\beq\label{syn}
	\dot r_c(\sc)=\dot r\en(\sen) \qquad (\text{simultaneity condition})
\enq
holds from which it follows that 
$$\ac\sc=\aen\sen.$$

 Then let us take the Lorentz boost (see \eqref{boost}) from $\uc$ to $\dot r_c(\sc)$,
\beq\label{cboost}
\Lt(\sc):=\1 + \frac{(\dot r_c(\sc) + \uc)\otimes(\dot r_c(\sc) + \uc)}{1-\dot r_c(\sc)\cdot\uc} - 2\dot r_c(\sc)\otimes\uc.
\enq
Using
\beq\label{ch1}
\dot r_c(\sc) + \uc= \uc\bigl(\cosh(\ac\sc) +1\bigr) \quad \text{and} \quad 1- \dot r_c(\ac\sc)=1+\cosh(\ac\sc),
\enq
a short calculation results in 
\beq\label{ln}
	L(\sc)\cdot\nn= \nn + (\nc\cdot\nn)\bigl(\uc\sinh(\ac\sc) + \nc(\cosh(\ac\sc-1)\bigr)
\enq
and 
\beq\label{lndot}
	\dot \Lt(\sc)\cdot\nn= (\nc\cdot\nn)\ac\bigl(\uc\cosh(\ac\sc) + \nc\sinh(\ac\sc)\bigr).
\enq

All these imply that the range of   
\beq\label{repl} 
	r_c(\sc) + \ep \Lt(\sc)\cdot\nn= x_c + \ep\nn + \bigl(\uc \sinh(\ac\sc) + \nc (\cosh(\ac\sc)-1)\bigr)\left(\frac{1}{\ac} + \ep \nc\cdot\nn\right)
\enq
equals the range of $\ren$; therefore we have
$$
	\aen=\frac{\ac}{1 + \ep\ac(\nc\cdot\nn)}, \quad 
	\sc=\frac{\sen}{1 + \ep\ac(\nc\cdot\nn)}.
$$

Moreover, we obtain
\beq\label{dotrdotl}
	\dot r_c(\sc) + \ep\dot \Lt(\sc)\cdot\nn= 
	(1+\ep\ac(\nc\cdot\nn))\dot r\en\left((1 + \ep\ac(\nc\cdot\nn)\sc)\right).
\enq

The sphere of radius $\ep$ with centre $\ran(r_c)$ in spacetime is the subset 
$$
	T_\ep:=\bigcup_{\nn\in S_c(1)}\ran(\ren),
$$
which is a world tube of type treated in \cite{BilEta19a}: 
the distance between  $\ran(r\en)$ and $\ran(r_c)$ equals $\ep$ at every synchronization instant.

\section{Lebesgue measure on the world tube}

For the notions and formulae appearing in this section, we refer to the subsections \ref{subman} and \ref{lebes} of the Appendix.

The world tube $T_\ep$ is a three dimenional submanifold in spacetime, its very definition 
gives the map
\beq\label{genpar}
	p:\I\times S_c(1)\to \aM,  \quad (\sc,\nn)\mapsto r_c(\sc) + \ep \Lt(\sc)\cdot\nn;
\enq
putting here $\nn:=p_c(\vartheta,\varphi)$, we get a parametrization of $T_\ep$; accordingly,
\eqref{genpar} is called a generalised parametrization and its use admits a concise formulation in the
following. The derivative of $p$ is
\beq\label{zz}
	\Dd p[\sc,\nn]= \begin{pmatrix} \partial_{\sc} p(\sc\nn) & \partial_\nn  p(\sc,\nn)\end{pmatrix} =
	\begin{pmatrix}\dot r_c(\sc) + \ep\dot \Lt(\sc)\cdot\nn & \ep \Lt(\sc)|_{\E_{\uc\nn}}\end{pmatrix}
\enq
where the last symbol denotes the restriction of the Lorentz boost to the linear subspace in question, $\E_{\uc\nn}$. 

For the sake of brevity, using the notation 
\beq\label{zzz}
	\z(\sc,\nn):=\dot r_c(\sc) + \ep\dot \Lt(\sc)\cdot\nn
\enq 
and then omitting the variables and subscripts, we have
\beq\label{dpdp}
(\Dd p)^*\cdot\Dd p= 
	\begin{pmatrix} \z \\ \ep \left(\Lt|_\E\right)^*\end{pmatrix}
	\begin{matrix}\cdot\begin{pmatrix} \z & \ep \Lt|_\E\end{pmatrix} \\
		 & & \end{matrix} = 
	\begin{pmatrix} \z\cdot\z & \z\cdot\ep \Lt|_\E\\
		\ep \left(\Lt|_\E\right)^*\cdot\z & \ep^2\1_\E\end{pmatrix}.
\enq

Note that the adjoint of $\Lt|_\E:\E\to\M$ is the linear map $(\Lt|_\E)^*:\M\to\E$ defined by $((\Lt|_\E)^*\cdot\x)\cdot\q=\x\cdot\Lt|_\E\cdot\q= \x\cdot\Lt\cdot\q$ for all $\x\in\M$, $\q\in\E$. 

Thus, recalling the projection $\Pp$ onto $\E$, for all vectors $\y$ we have, 
$$
	((\Lt|_\E)^*\cdot\z)\cdot(\Pp\cdot\y) =
	\z\cdot\Lt\cdot(\Pp\cdot\y) =
	(\Lt^*\cdot\z)\cdot(\Pp\cdot\y) = 
	(\Pp\cdot\Lt^*\cdot\z)\cdot\y
$$ 
which means that the block matrix form of $(\Dd p)^*\cdot\Dd p$ is symmetric:
$\z\cdot\Lt|_\E=\left(\Lt|_\E\right)^*\cdot\z=\Pp\cdot\Lt^*\cdot\z\in\E$. 
                       
As a consequence, since $\E$ is two dimensional\footnote{The matrix \eqref{dpdp} in spherical coordinates has the form 
$\begin{pmatrix} \alpha & \ep\beta_1 & \ep\beta_2\\
\ep\beta_1 & \ep^2 & 0 \\ \ep\beta_2 & 0  &\ep^2\end{pmatrix}$}
\beq\label{detdpdp}
	\det((\Dd p)^*\cdot\Dd p) = 
	\ep^4\z\cdot\z - \ep^4\left(\Pp\cdot\Lt^*\cdot\z\right)\cdot\left(\Pp\cdot\Lt^*\cdot\z\right).
\enq

Here
\begin{align*}
	\left(\Pp\cdot\Lt^*\cdot\z\right)\cdot\left(\Pp\cdot\Lt^*\cdot\z\right)&=
	\left(\Lt^*\cdot\z\right)\cdot(\1 + \uc\otimes\uc - 
		\nn\otimes\nn)\cdot\Lt^*\cdot\z=\\ &= 
	\z\cdot\z + (\z\cdot \Lt\cdot\uc)^2 - (\z\cdot \Lt\cdot\nn)^2.
\end{align*}

Equalities $\dot r_c\cdot\Lt\cdot\nn=0$ and $(\dot\Lt\cdot\nn)\cdot(\Lt\cdot\nn)=0$, together with the definition 
\eqref{zzz} imply that $\z\cdot\Lt\cdot\nn=0$.

Further, \eqref{dotrc}  and \eqref{lndot} imply $(\dot\Lt\cdot\nn)\cdot\dot r_c=-\ac(\nc\cdot\nn)$, thus from \eqref{zzz} and the property $\Lt\cdot\uc=\dot r_c$ of the Lorentz boost we have that $\z\cdot\Lt\cdot\uc=-(1+\ep\ac(\nc\cdot\nn))$.

Finally, according to \eqref{dotrdotl}, $\z\cdot\z=-\bigl(1 + \ep\ac(\nc\cdot\nn)\bigr)^2$. 

Summing up, the second term of \eqref{detdpdp} is zero and 
$$
	\sqrt{|\det{\Dd p^*[\sc,\nn]\cdot\Dd p[\sc,\nn]}|}=
		\ep^2\bigl(1 + \ep\ac(\nc\cdot\nn)\bigr),
$$
and the Lebesgue measure $\la_{T_\ep}$ on the tube is given by the integration formula
\beq\label{late}
	\int f\ d\la_{T_\ep}=
		\ep^2\int_{\I}\int_{S_c(1)} f\bigl(r_c(\sc) + \ep L(\sc)\nn)\bigr)
		\bigl(1 + \ep\ac(\nc\cdot\nn)\bigr)\ d\nn \ d\sc.
\enq

\section{Uniform charge density on the sphere -- \\ world current}

According to \eqref{syn} and \eqref{lndot}, the set of spacetime points simultaneous with $r_c(\sc)$ is the hyperplane (in fact only a convenient part of it) which contains $r_c(\sc)$ and is Lorentz-orthogonal to $\dot r_c(\sc)$. Thus, at every synchronization instant the world tube $T_\ep$ (i.e. the intersection of the tube and the instant hyperplane) is a sphere of radius $\ep$. The world current of a uniform charge density $\sigma$ on the sphere is the vector measure 
$$
	\jj_\ep:=\sigma\uu_\ep\la_{T_\ep}
$$ 
where $\uu_\ep$ is the function (a vector field) on the world tube which assigns the corresponding absolute velocities to the points, 
i.e. $\uu_\ep(r_{\ep\nn}):=\dot r_{\ep,\nn}$, in other words, by \eqref{dotrdotl},
\beq\label{tue}
	\uu_\ep\bigl(r_c(\sc) + \ep\Lt(\sc)\cdot\nn\bigr)=
	\frac{\dot r_c(\sc) + \ep \dot\Lt(\sc)\cdot\nn}{1 + \ep\ac(\nc\cdot\nn)}.
\enq

\section{Electromagnetic potentials}

For the notions and formulae appearing in this section, we refer to the subsection \ref{distr} of the Appendix. 

From now on, for the sake of brevity, we write $s$ and $a$ instead of $\sc$ and $\ac$. 

Now we are ready to compare the electromagnetic potentials of a point charge and the one of a charged spherical shell. In electrostatics, they are equal outside the shell; however, we will show that they are different in case of uniform acceleration.

The electromagnetic potential produced by the uniformly accelarated charged sphere is the distribution $\jj_\ep*\la_{\Ll}$ which, according to \eqref{late} and \eqref{tue}, acts on a test function $\Phi$ as follows:
\begin{align}
	&(\jj_\ep*\la_{\Ll}\mid\Phi)=\int\int \Phi(x + \x)\ 
		\sigma \uu_\ep(x)\ d\la_{T_\ep}(x)\ d\la_\Ll(\x)=\nonumber\\ 
	&=\sigma\ep^2\int_{\I} \int_{S_c(1)}\int_{\Ll}  
		\Phi(r_c(s) + \ep \Lt(s)\cdot\nn + \x)\bigl(\dot r_c(s) + \ep\dot \Lt(s)\cdot\nn)\bigr)\ d\nn\ ds \ d\la_{\Ll}(\x). 
\label{csopot}
\end{align}

Let us consider now a point charge $\sigma 4\pi\ep^2$ existing on the centre of the world tube. The corresponding world current is
$$
	\jj_c:=\sigma 4\pi\ep^2 \dot r_c\la_{\ran r_c}
$$
producing the electromagnetic potential $\jj_c*\la_{\Ll}$ for which\footnote{The known actual form of the potential is obtained by the substitution $x:=r_c(s) + \x$ from which the retarded proper time $s_{\text{ret}}(x)$ is determined in such a way that $x-r(s_{\text{ret}}(x))$ be lightlike}
\beq\label{pontpot}
	(\jj_c*\la_{\Ll}\mid\Phi)=
		\sigma 4\pi\ep^2\int_{\Ll}\int_{\I}\Phi(r_c(s) + \x)\dot r_c(s)\ ds \ d\la_{\Ll}(\x).
\enq

{\it The electromagnetic potentials produced by the uniformly accelerated charged sphere and by the point charge would be equal outside the world tube if and only if the integrals \eqref{csopot} and \eqref{pontpot} were equal for all $\Phi$ having support outside the world tube}. 

Taking such a $\Phi$ and rewriting \eqref{pontpot} in the form
\beq\label{pontpotuj}
	\sigma \ep^2 \int_{\Ll}\int_{\I}\int_{S_c(1)}\Phi(r_c(s) + \x)\dot r_c(s) \ d\nn \ ds \ d\la_{\Ll}(\x),
\enq
let us examine the difference of the integrals.  

According to the equality
\begin{align*}
	&\Phi\big(r_c(s) + \ep \Lt(s)\!\cdot\!\nn + \x\bigr)=\Phi\bigl(r_c(s) +\x\bigr) +\\
	&\quad +(\ep\Lt(s)\!\cdot\!\nn)\!\cdot\!\Dd\Phi[r_c(s)+\x] +(\ep \Lt(s)\!\cdot\!\nn)\!\cdot\!\Dd^2\Phi[r_c(s) +\x]\!\cdot\!(\ep \Lt(s)\cdot\nn) 
+ \ordo(\ep^2),
\end{align*} 
the difference of the integrands, omitting the variables for the sake of perspicuity, becomes                           
\begin{multline} 
	\Bigl(\Phi + (\ep \Lt\cdot\nn)\cdot\Dd\Phi + 
		(\ep\Lt\cdot\nn)\cdot\Dd^2\Phi\cdot(\ep\Lt\cdot\nn)\Bigr) \ep\dot\Lt\cdot\nn  +\\
	\Bigl((\ep \Lt\cdot\nn)\cdot\Dd\Phi + (\ep \Lt\cdot\nn)\cdot\Dd^2\Phi\cdot
		(\ep\Lt\cdot\nn)\Bigr)\dot r_c + \ordo(\ep^2).
\end{multline}

The integration by $\nn$ (see \eqref{nnn}) yields zero for the terms linear and trilinear in $\nn$. As concerns the bilinear 
terms, $\bigl((\Lt\cdot\nn)\cdot\Dd\Phi\bigr)\dot\Lt\cdot\nn =\dot\Lt\cdot(\nn\otimes\nn)\cdot\Lt^*\cdot\Dd\Phi$ and
$(\Lt\cdot\nn)\cdot\Dd^2\Phi\cdot(\Lt\cdot\nn)=\nn\cdot\bigl(\Lt^*\cdot\Dd\Phi\Lt\bigr)\nn$; consequntly, the integration by $\nn$ 
yields (recall that $\Lt^*=\Lt^{-1}$ and a cyclic permutation can be made under a trace)
\beq\label{kettag}
\ep^2\frac{4\pi}{3}\Bigl(\dot\Lt\cdot(1+\uc\otimes\uc)\cdot\Lt^*\cdot\Dd\Phi + \Tr\left(\nabla_{\uc}^2\Phi\right)\dot r_c\Bigr) .
\enq

To find the properties of the first term, we proceed as follows. Since $\uc\cdot\Lt^*=\Lt\cdot\uc=\dot r_c$, we have 
\beq
\label{ezaz}\dot\Lt\cdot(1+\uc\otimes\uc)\cdot\Lt^*=\dot\Lt\cdot\Lt^* + \ddot r_c\otimes\dot r_c.
\enq

$\dot\Lt\cdot\Lt^*$ is evidently the linear combination of $\uc\otimes\uc$, $\uc\otimes\nc$, $\nc\otimes\uc$ and $\nc\otimes\nc$; 
the coefficients are obtained by $\uc\cdot\dot\Lt\cdot\Lt^*\cdot\uc=(\uc\cdot\dot\Lt)\cdot(\uc\cdot\Lt)$ etc.

A simple calculation based on \eqref{cboost} and \eqref{ch1} yields
$$ \uc\cdot\Lt(s)=\uc\cosh(as) - \nc\sinh(as) \qquad \uc\cdot\dot\Lt(s)=a\bigl(\uc\sinh(as) -\nc\cosh(as)\bigr), $$
$$\nc\cdot\Lt(s)=-\uc\sinh(as) + \nc\cosh(as) \qquad \nc\cdot\dot\Lt(s)=a\bigl(-\uc\cosh(as) +\nc\sinh(as)\bigr) $$
from which it follows that $\dot\Lt\cdot\Lt^*=a(\uc\otimes\nc - \nc\otimes\uc)$. At last, 
\begin{multline*}
a(\uc\otimes\nc - \nc\otimes\uc) + a(\uc\sinh + \nc\cosh)\otimes(\uc\cosh + \nc\sinh)= \\
a(\uc\cosh + \nc\sinh)\otimes(\uc\sinh +\nc\cosh) = \dot r_c\otimes\ddot r_c
\end{multline*}
and we have that the first term in \eqref{kettag} is
$$
\Bigl(\ddot r_c(s)\cdot\Dd\Phi[r_c(s) + \x]\Bigr)\dot r_c(s).
$$

As a consequence, we can state that the two electromagnetic potentials would be equal if and only if  
$$\int_\I\int_{\Ll}\Bigl(\ddot r_c(s)\cdot\Dd\Phi[r_c(s) + \x] + 
\Tr\left(\nabla_{\uc}^2\Phi[r_c(s)+ \x]\right)\Bigr)\dot r_c(s)\ ds\ d\la_\Ll(\x)$$
were zero for all $\Phi$ having support outside the world tube. The integrand has timelike values or zero, it is evident then that the integral cannot be zero for all $\Phi$ in question.

As a consequence, {\it the electromagnetic potentials produced by the uniformly accelarated charged sphere and by the point charge at the centre are not equal outside the world tube}.

\section{Conclusions}

The physical and mathematical paradoxes of LAD equation challenge the foundations of physics since more than a century. It is widely believed that the problem is not physical and the instabilities are due to the oversimplified 
mathematical model of point charge. Therefore a continuum theory, a charge distribution is considered a better model that eventually can remove the paradoxes. 

The simplest possible continuum model is the uniformly charged rigid spherical shell, a model that removes the singularity and domain incompatibility of the Newton-Maxwell system of equations. It is continuously kept analysed and improved since the seminal work of Lorentz \cite{Lor892a}.
{\textcolor{blue} In this work we have shown that in case of accelerated motion a homogenous charge distributions the shell cannot be substituted by any central charge, contrary to our electrostatic based intuition.



When treating the relation of point charges and continuous charge distributions one tempted to keep the universality introducing arbitrary shapes with minimal symmetry requirements. As it is stated in the Introduction, the theory of distributions (generalized functions) seems to be the proper tool for an adequate mathematical treatment of point charges, because in this framework the singularities can be ruled out. Spherical shell charge distributions are convenient and calculable therefore they can be good models to test more general approaches, when point charges are approximated by charged continua.

\section{Appendix}

\subsection{Tensor products}\label{tensor} 

\quad \ \ 1. The tensor product $\ad\otimes\bd$ of the vectors $\ad$ and $\bd$ can be considered either a linear or a bilinear map; its action on the vector $\x$ or the vectors $\y$ and $\x$ is $(\ad\otimes\bd)\cdot\x:=\ad(\bd\cdot\x)$ or, $\y\cdot(\ad\otimes\bd)\cdot\x:=(\y\cdot\ad)\bd\cdot\x)$.

2. For an absolute velocity $\uu$, $\1 +\uu\otimes\uu$ is the projection onto the linear subspace $\Sd_\uu$ of $\uu$-spacelike vectors.

3. For two absolute velocities $\uu$ and $\uu'$, the Lorentz boost from $\uu$ to $\uu'$ is 
\beq\label{boost}
	\1 + \frac{(\uu' + \uu)\otimes(\uu' + \uu)}{1-\uu'\cdot\uu} -2\uu'\otimes\uu,
\enq 
which sends $\uu$ to $\uu'$ and maps the Euclidean subspace $\Sd_\uu$ onto the Euclidean subspace $\Sd_{\uu'}$ in a rotation-free way.

\subsection{Derivatives}\label{deriv}

For more details, see section VI.3 of \cite{Mat20b}.

The symbol $\ordo$ means a function $\rr^+\to\rr$, defined in a neighbourhood of zero, such that $\lim_{\alpha\to0}\frac{\ordo(\alpha)}{\alpha}=0$.

The derivative of a differentiable function $\Phi:\aM\to\rr$ at $x$ is the linear map $\Dd\Phi[x]:\M\to\rr$ for which 
$$
	\Phi(x+ \x) = \Phi(x)+\x\cdot\Dd\Phi[x] + \ordo(\|\x\|) \qquad (\x\in\M)
$$
holds for an arbitrary norm $\| \ \|$ on $\M$.

The second derivative of a twice differentiable function $\Phi:\aM\to\rr$ at $x$ is the bilinear $\Dd^2\Phi[x]$ map for which 
$$
	\Phi(x+ \x) = \Phi(x)+\x\cdot\Dd\Phi[x] + \x\cdot\Dd^2\Phi[x]\cdot\x + \ordo(\|\x\|^2) \qquad (\x\in\M)
$$
holds.

Similar formulae are valid for functions defined in an arbitrary affine space and having values in $\aM$.

For an absolute velocity $\uu$, the $\uu$-spacelike derivative of $\Phi$ is the restriction of $\Dd\Phi[x]$ onto the linear subspace of $\uu$-spacelike vectors, 
$$
	\nabla_\uu\Dd\Phi[x]:=\Dd\Phi[x]|_{\Sd_\uu}=(\1 +\uu\otimes\uu)\cdot\Dd\Phi[x].
$$

\subsection{Submanifolds}\label{subman}

For more details, see section VI.4 of \cite{Mat20b}.

1. For $d\in\{0,1,2,3,4\}$ a submanifold $H$ of dimension $d$ in $\aM$ ($\M$) is a subset for which there is a smooth map $p:\rr^d\to\aM \ (\M)$, 
called a parametrization, such that 

-- \ $\ran(p)=H$,

-- \ $p$ is injective and $p^{-1}$ is continuous,

-- \ $\Dd p[\xi]:\rr^d\to\M$ is injective for all $\xi$ in the domain of $p$.

More generally, the domain of a parametrization can be a $d$ dimensional affine space instead of $\rr^d$.

The tangent space of $H$ over $p(\xi)$ is the range of $\Dd p[\xi]$, a linear subspace of $\M$.

2. For an absolute velocity $\uc$, the unit sphere $S_c(1)$ of $\uc$-spacelike vectors is a two dimensional submanifold which can be parametrized by the usual angles $\vartheta$ and $\varphi$: taking two unit vectors $\ad$ and $\bd$, orthogonal to each other and to $\nc$, 
\beq\label{s1par}
	p_c(\vartheta,\varphi):=\uc + \nc\cos\vartheta + \ad\sin\vartheta\cos\varphi + \bd\sin\vartheta\sin\varphi.
\enq

The tangent space of $S_c(1)$ over $\nn$ is the linear subspace $\E_{\uc\nn}$  consisting of vectors Lorentz orthogonal to both $\uc$ and $\nn$; it is the range of the linear projection $\Pp_{\uc\nn}:=(\1 + \uc\otimes\uc -\nn\otimes \nn)$.

3. The futurelike light cone 
$$
	\Ll:=\{\x\mid \x\cdot\x=0, \ \uu\cdot\x<0 \ \text{for all absolute velocities} \ \uu\}
$$
is a three dimensional submanifold in $\M$ which can be parametrized with the aid of an arbitrary absolute velocity $\uu$:
$$
	p(\q):= \uu|\q| + \q \qquad (\uu\cdot\q=0) \qquad (\q\in\Sd_\uu).
$$

\subsection{Lebesgue measures}\label{lebes}

For measure theory we refer to \cite{Din67b,AmaEsche09b}.

1. The Lebesgue measure on $\aM$ (on $\M$) is the translation invariant measure $\la_\aM$ ($\la_\M$) which assigns the value $|\ad|\ |\bd|\ |\mathbk c|\ |\mathbk d|$ to a prism defined by Lorentz orthogonal vectors $\ad,\bd,\mathbk c,\mathbk d$, where $| \ |$ denotes the pseudo-length.

For the integrals of functions $f$ defined in $\aM$, we use the notations
$$
	\int f\ d\la_\aM = \int_\aM f(x)\ dx.
$$
 
2. Let $H$ be a $d$ dimensional submanifold in $\aM$ (in $\M$). Then for a parametri\-zation $p$ of $H$, $(\Dd p[\xi])^*$ is a linear map $\M\to\rr^d$, thus $(\Dd p[\xi])^*\cdot\Dd p[\xi]$ is a linear map $\rr^d\to\rr^d$, i.e. it is a $d$ times $d$ matrix.

The Lebesgue measure $\la_H$ on the submanifold $H$ is defined in such a way that for a function $f$ defined in $H$, 
$$
	\int f \ d\la_H:= \int_{\rr^d} f(p(\xi))\sqrt{|\det\bigl((\Dd p[\xi])^*\cdot\Dd p[\xi]\bigr)}\ d\xi
$$
for an arbitrary parametrization $p$ (thus, the integral is the same for all $p$).

Of course, also a function defined in $\aM$ ($\M$) can be integrated by $\la_H$, taking the restriction of $f$ onto $H$; in this way we can -- and we do -- consider $\la_H$ a measure on $\aM$ ($\M$), too.

If $\h:H\to\M$ is a continuous function then $\h\la_H$ is a vector measure, for which the integrals are defined by $d(\h\la_H):=\h\ d\la_H$.

3. The Lebesgue measure on the unit sphere, $S_c(1)$, is given by the well known formula
$$
	\int f \ d\la_{S_c(1)}= \int_{S_c(1)} f(\nn)\ d\nn= 
		\int_0^{2\pi}\int_0^\pi f(p(\vartheta,\varphi))\ \sin\vartheta\ d\vartheta\ d\varphi.
$$ 

Then it is easy to find that
\beq\label{nnn}
	\int_{S_c(1)} \nn \ d\nn =0,  \quad
	\int_{S_c(1)} \nn\otimes\nn \ d\nn=\frac{4\pi}{3}(\1 + \uc\otimes\uc), \quad
	\int_{S_c(1)} \nn\otimes\nn\otimes\nn \ d\nn = 0,
\enq
which go over tensors, too; for instance,
\begin{align*}
	\int_{S_c(1)} \nn\cdot(\ad\otimes\bd)\cdot\nn\ d\nn 
		=&\int_{S_c(1)} \ad\cdot(\nn\otimes\nn)\cdot\bd\ d\nn =\\
		=&\frac{4\pi}{3}\ad\cdot(\1 + \uc\otimes\uc)\cdot\bd=\\
		=& \frac{4\pi}{3}\bigl((\1+\uu\otimes\uc)\cdot\ad\bigr)(\bd\cdot(\1+\uu\otimes\uu)\bigr)=\\
		=&\label{nten}\frac{4\pi}{3}\Tr\Bigl((\1 + \uc\otimes\uc)\cdot(\ad\otimes\bd)\cdot(\1 + \uc\otimes\uc)\Bigr),
\end{align*}
and the last formula holds for arbitrary bilinear maps instead of $\ad\otimes\bd$.

4. According to the general definition, the Lebesgue measure on the futurelike light cone, $\Ll$, is zero. Nevertheless, we can define on it a nonzero measure quite naturally, which we consider to be the Lebesgue measure. Namely, for all $a>0$ 
$$
	V(a):=\{\x\in\M\mid \x\cdot\x=-a^2, \ \x \ \text{is futurelike}\}
$$
is a three dimensional submanifold which can be parametrized with the aid of an arbitrary absolute velocity $\uu$:
$$
	p(\q):= \uu\sqrt{|\q|^2 +a^2} + \q \qquad (\q\in\Sd_\uu).
$$
It is simply obtained that $\sqrt{|\det(\Dd p[\q]^*\cdot\Dd p[\q])|}=\frac{a}{\sqrt{|\q|^2 + a^2}}$. Then the Lebesgue measure on the light cone is defined to be 
$$
	\la_{\Ll}:=\lim_{a\to 0}\frac1{a}\la_{V(a)}
$$
in an appropriate sense of the limit procedure; then we have the integration formula 
$$
	\int f\ d\la_{\Ll}= \int_{\Ll} f(\x)\ d\x = \int_{\Sd_\uu} f(\uu|\q| +\q)\ \frac{d\q}{|\q|}
$$
for an arbitrary absolute velocity $\uu$.

\subsection{Distributions}\label{distr}

For distribution theory we refer to \cite{Hor12b,Dij13b}.

The usual setting of distribution theory is based on $\rr^n$. It is a quite simple generalisation that we consider the affine space $\aM$ instead of $\rr^n$. Thus, the space of our test functions, $\Dt(\aM)$, consists of smooth functions $\Phi:\aM\to\rr$, with compact support. A distribution is a continuous linear map $\Dt(\aM)\to\rr$.

Another simple generalisation that we consider vector distributions, too, i.e. continuous linear maps $\Dt(\aM)\to\M$. The action of a (vector) distribution $T$ on the test function $\Phi$ is denoted by $(T\mid\Phi)$.

The present article needs only the following facts from distribution theory. 

1. The derivative of a (vector) distribution $T$ is the distribution $\Dd T$ defined by
$$
	(\ad\cdot\Dd T\mid\Phi):=-(T\mid \ad\cdot\Dd\Phi) \qquad (\ad\in\M).
$$

2. A Lebesgue measure of a submanifold $H$ is a distribution by the definition
$$
	(\la_H\mid\Phi):=\int \Phi\ d\la_H =\int_H \Phi(x)\ dx.
$$

3. If $H$ is a submanifold in $\aM$, $f$ is a continuous function defined in $H$,  and $K$ is a submanifold in $\M$, the convolution $f\la_H*\la_K$ is the 
distribution defined by
$$
	(f\la_H*\la_K\mid\Phi):=\int_H\int_K f(x)\Phi(x + \x)\ dx\ d\x.
$$

4. For the d'Alembert operator $\square:=\Dd\cdot\Dd:=\Tr(\Dd^2)$ (in coordinates $\partial_k\partial^k$)
$$
	\square(T*\la_{\Ll})=T.
$$

\section{Acknowledgement}   

The work was supported by the grants National Research, Development and Innovation Office NKFIH 124366(124508) and NKFIH 123815. The support of TKP is acknowledged. 

This work was performed in the frame of the FIEK 16-1-2016-0007 project, implemented with the support provided from the National Research, Development and Innovation Fund of Hungary, financed under the FIEK 16 funding scheme. Some parts of the research reported in this paper have been supported by the National Research, Development and Innovation Fund (TUDFO/51757/2019-ITM), Thematic Excellence Program."


\begin{thebibliography}{10}
	
	\bibitem{Lor892a}
	H.A. Lorentz.
	\newblock La th{\'e}orie {\'e}lectromagn{\'e}tique de {M}axwell et son
	application aux corps mouvants.
	\newblock {\em Archives Neerlandaises des Sciences Exactes et Naturelles},
	25:363--552, 1892.
	
	\bibitem{Abr905b}
	Max Abraham.
	\newblock {\em Theorie der {E}lektrizit{\"a}t, {Vol. II:} Elektromagnetische
		{T}heorie der {S}trahlung}.
	\newblock Teubner, Leipzig, 1905.
	
	\bibitem{Lor09b}
	H.A. Lorentz.
	\newblock {\em The Theory of Electrons}.
	\newblock Teubner, Leipzig, 1909.
	
	\bibitem{Dir38a}
	P.~A.~M. Dirac.
	\newblock Classical theory of radiating electrons.
	\newblock {\em Proceedings of the Royal Society of London. Series A.
		Mathematical and Physical Sciences}, 167(929):148--169, 1938.
	
	\bibitem{BorInf34a}
	M.~Born and L.~Infeld.
	\newblock Foundations of the new field theory.
	\newblock {\em Proceedings of the Royal Society of London. Series A},
	144(852):425--451, 1934.
	
	\bibitem{Bop40a}
	F.~Bopp.
	\newblock Eine lineare {T}heorie des {E}lektrons.
	\newblock {\em Annalen der Physik}, 430(5):345--384, 1940.
	
	\bibitem{Kie19a}
	M.~K.-H. Kiessling.
	\newblock Force on a point charge source of the classical electromagnetic
	field.
	\newblock {\em Physical Review D}, 100(6):065012, 2019.
	\newblock arXiv:1907.11239v4, with erratum.
	
	\bibitem{ForOco91a}
	G.W. Ford and O'Connell.
	\newblock Radiation reaction in electrodynamics and the elimination of runaway
	solutions.
	\newblock {\em Physics Letters A}, 157(4-5):217--220, 1991.
	
	\bibitem{Ver08a}
	J.~Verh\'as.
	\newblock White areas on the map of applying non-equilibrium thermodynamics: on
	the self accelerating electron.
	\newblock {\em Atti dell'Accademia Peloritana dei Pericolanti, Classe di
		Scienze Fisiche, Matematiche e Naturali, Suppl. I.}, 86:1--7, 2008.
	
	\bibitem{Pol14a}
	J.~Polonyi.
	\newblock Effective dynamics of a classical point charge.
	\newblock {\em Annals of Physics}, 342:239--263, 2014.
	
	\bibitem{ForEta20a}
	M.~Formanek, A.~Steinmetz, and J.~Rafelski.
	\newblock Radiation reaction friction: {R}esistive material medium.
	\newblock {\em Physical Review D}, 102:056015, 2020.
	\newblock arXiv:2004.09634.
	
	\bibitem{Spo00a}
	H.~Spohn.
	\newblock The critical manifold of the {Lorentz-Dirac} equation.
	\newblock {\em EPL (Europhysics Letters)}, 50(3):287, 2000.
	
	\bibitem{MatEta17a}
	Matolcsi T., F\"ul\"op T., and Weiner M.
	\newblock Second order equation of motion for electromagnetic radiation
	back-reaction.
	\newblock {\em Journal of Modern Physics}, 32(27):1750147(18), 2017.
	\newblock arXiv:1207.0428.
	
	\bibitem{Par87b}
	S.~Parrott.
	\newblock {\em Relativistic electrodynamics and differential geometry}.
	\newblock Springer, 2012.
	
	\bibitem{BilEta19a}
	C.~Bild, D.-A. Deckert, and H.~Ruhl.
	\newblock Radiation reaction in classical electrodynamics.
	\newblock {\em Physical Review D}, 99(9):096001, 2019.
	
	\bibitem{MonSha74a}
	E.~J. Moniz and D.H. Sharp.
	\newblock Absence of runaways and divergent self-mass in nonrelativistic
	quantum electrodynamics.
	\newblock {\em Physical Review D}, 10(4):1133, 1974.
	
	\bibitem{MonSha77a}
	E.~J. Moniz and D.H. Sharp.
	\newblock Radiation reaction in nonrelativistic quantum electrodynamics.
	\newblock {\em Physical Review D}, 15(10):2850, 1977.
	
	\bibitem{GalEta10a}
	C.~R. Galley, A.~K. Leibovich, and I.~Z. Rothstein.
	\newblock Finite size corrections to the radiation reaction force in classical
	electrodynamics.
	\newblock {\em Physical review letters}, 105(9):094802, 2010.
	
	\bibitem{ForEta12c}
	P.~Forg{\'a}cs, T.~Herpay, and P.~Kov{\'a}cs.
	\newblock Comment on "{Finite Size Corrections to the Radiation Reaction Force
		in Classical Electrodynamics}".
	\newblock {\em Physical review letters}, 109(2):029501, 2012.
	
	\bibitem{GalEta12c}
	C.~R. Galley, A.K. Leibovich, and I.Z. Rothstein.
	\newblock Galley, {L}eibovich, and {R}othstein reply.
	\newblock {\em Physical Review Letters}, 109(2):029502, 2012.
	
	\bibitem{Kos19a}
	B.P. Kosyakov.
	\newblock Self-interaction in classical gauge theories and gravitation.
	\newblock {\em Physics Reports}, 812:1--55, 2019.
	
	\bibitem{Pol19a}
	J.~Polonyi.
	\newblock The {Abraham--L}orentz force and electrodynamics at the classical
	electron radius.
	\newblock {\em International Journal of Modern Physics A}, 34(15):1950077,
	2019.
	
	\bibitem{Spo04b}
	H.~Spohn.
	\newblock {\em Dynamics of charged particles and their radiation field}.
	\newblock Cambridge university press, 2004.
	
	\bibitem{Yag06b}
	Arthur Yaghjian.
	\newblock {\em Relativistic dynamics of a charged sphere: updating the
		{L}orentz-{A}braham model}.
	\newblock Springer, 2006.
	\newblock 2nd edition, first: 1992.
	
	\bibitem{Roh07b}
	Fritz Rohrlich.
	\newblock {\em Classical charged particles}.
	\newblock World Scientific Publishing Company, 2007.
	\newblock 3rd edition, original: 1965.
	
	\bibitem{Kos07b}
	Boris Kosyakov.
	\newblock {\em Introduction to the classical theory of particles and fields}.
	\newblock Springer, 2007.
	
	\bibitem{WisEta18a}
	T.N. Wistisen, A.~Di~Piazza, H.~V. Knudsen, and U.~I. Uggerh${\o}$j.
	\newblock Experimental evidence of quantum radiation reaction in aligned
	crystals.
	\newblock {\em Nature communications}, 9(1):1--6, 2018.
	
	\bibitem{PodEta18a}
	K.~Poder, M.~Tamburini, G.~Sarri, A.~Di~Piazza, S.~Kuschel, C.D. Baird,
	K.~Behm, S.~Bohlen, J.M. Cole, D.J. Corvan, and +other 14~authors.
	\newblock Experimental signatures of the quantum nature of radiation reaction
	in the field of an ultraintense laser.
	\newblock {\em Physical Review X}, 8(3):031004, 2018.
	
	\bibitem{Bla20a}
	T.G. Blackburn.
	\newblock Radiation reaction in electron--beam interactions with high-intensity
	lasers.
	\newblock {\em Reviews of Modern Plasma Physics}, 4(1):1--37, 2020.
	
	\bibitem{LanLif78b}
	L.~D. Landau and E.~M. Lifshitz.
	\newblock {\em The Classical Theory of Fields (Course of Theoretical Physics,
		vol. 2)}.
	\newblock Pergamon Press, Oxford, 2th edition, 1959.
	
	\bibitem{Sok09a}
	I.V. Sokolov.
	\newblock Renormalization of the {Lorentz-Abraham-Dirac} equation for radiation
	reaction force in classical electrodynamics.
	\newblock {\em Journal of Experimental and Theoretical Physics},
	109(2):207--212, 2009.
	
	\bibitem{Zot16a}
	D.B. Zot'ev.
	\newblock Critical remarks on {S}okolov's equation of the dynamics of a
	radiating electron.
	\newblock {\em Physics of Plasmas}, 23(9):093302, 2016.
	
	\bibitem{Ger96m}
	R.~Geroch.
	\newblock Partial differential equations of physics.
	\newblock arXiv:gr-qc/9602055.
	
	\bibitem{GraEta09a}
	S.E. Gralla, A.I. Harte, and R.M. Wald.
	\newblock Rigorous derivation of electromagnetic self-force.
	\newblock {\em Physical Review D}, 80(2):024031, 2009.
	
	\bibitem{Dij13b}
	G.~van Dijk.
	\newblock {\em Distribution theory: convolution, {F}ourier transform, and
		{L}aplace transform}.
	\newblock Walter de Gruyter, 2013.
	
	\bibitem{Mat20bh}
	T.~Matolcsi.
	\newblock {\em Electromagnetism in spacetime}.
	\newblock 2020.
	\newblock to appear in Hungarian.
	
	\bibitem{Mat20m}
	T.~Matolcsi.
	\newblock On the Radiation Reaction Force.
	\newblock arXiv:2102.04983.
	
	\bibitem{RajRaj08a}
	S. Raju and C.K. Raju.
	\newblock Radiative damping and functional differential equations.
	\newblock {\em odern Physics Letters A}, 26(35):2627, 2011.
	\newblock arXiv:0802.3390.
	
	\bibitem{Mat20b}
	T.~Matolcsi.
	\newblock {\em Spacetime Without Reference Frames}.
	\newblock Minkowski Institute Press, 2 edition, 2020.
	
	\bibitem{Din67b}
	N.~Dinculeanu.
	\newblock {\em Vector measures}.
	\newblock Pergamon Press, 1967.
	
	\bibitem{AmaEsche09b}
	H.~Amann and J.~Escher.
	\newblock {\em Analysis {III.}}
	\newblock Birkh\"auser, 2009.
	
	\bibitem{Hor12b}
	Horv{\'a}th J.
	\newblock {\em Topological vector spaces and distributions}.
	\newblock Dover, 2012.
	
\end{thebibliography}

\end{document}